\documentclass[twocolumn,notitlepage,preprintnumbers,showpacs,showkeys,secnumarabic,amssymb,amsmath,amsfonts,prl]{revtex4-1}

\usepackage[T1]{fontenc}
\usepackage[latin9]{inputenc}
\usepackage{amsmath}
\usepackage{graphicx}
\usepackage{amssymb}
\usepackage{esint}

\begin{document}

\title{Optical Rogue Waves in Vortex Turbulence}

\author{Christopher J. Gibson, Alison M. Yao, and Gian-Luca Oppo}

\affiliation{SUPA and Department of Physics, University of Strathclyde, Glasgow
G4 0NG, Scotland, U.K.}

\begin{abstract}
We present a spatio-temporal mechanism for producing 2D optical rogue waves in the presence of a turbulent state with creation, interaction and annihilation of optical vortices. Spatially periodic structures with bound phase lose stability to phase unbound turbulent states in complex Ginzburg-Landau and Swift-Hohenberg models with external driving. When the pumping is high and the external driving is low, synchronized oscillations are unstable and lead to spatio-temporal {\it vortex-mediated} turbulence with high excursions in amplitude. Nonlinear amplification leads to rogue waves close to turbulent optical vortices, where the amplitude tends to zero, and to probability density functions (PDFs) with long tails typical of extreme optical events.
\end{abstract}

\maketitle
Rogue waves (RWs) in high seas were once a thing of legend: massive walls of water tens of metres high, capable of destroying large ships, appearing from nowhere then vanishing leaving no trace of their existence \cite{RW1}. Because of the very short lifetime of such extreme events, quantitative studies and simulations of the mechanisms behind their creation in oceanography have grown only recently \cite{Onorato13,Abrashkin13}. Although the origin of these waves is still under debate, RWs have been realized in several optical systems \cite{Dudley14} from optical fibres \cite{Solli07}, to optical cavities \cite{Montina09,Bonatto11} and photonic crystals \cite{Liu15}.

Here we investigate a spatio-temporal mechanism involving vortices in a 2D turbulent state that is capable of generating RWs, building upon previous work concerning a singly resonant optical parametric oscillator system with a low amplitude detuned seeding field \cite{Oppo13}. For generality and application to a variety of nonlinear dynamical systems, we consider a Forced Complex Ginzburg-Landau (FCGL) equation \cite{Coullet92} and a Forced Complex Swift-Hohenberg (FCSH) equation, both with external driving. We focus on the loss of synchronisation of the Adler locked states obtained at large driving amplitudes. When decreasing the external forcing, oscillations at the Adler frequency become spatio-temporally unstable leading first to a phase and then to an amplitude instability that forces, locally and randomly, the formation of pairs of oppositely charged vortices. Since the total power in the transverse direction remains almost constant throughout, the nonlinearity pushes the intensity to high spikes close to interacting vortices, resulting in the rare formation of RWs. The RWs described here are outside thermodynamic equilibrium, do not survive in the purely temporal (single mode) case and are due to a deterministic, nonlinear and vortex-mediated turbulence far removed from a purely stochastic superposition of optical waves.

To demonstrate the generality of optical RWs in vortex turbulence we employ a variety of mathematical models:
\small
\begin{eqnarray}
\label{model}
\partial_t E &=& E_{IN}-(1-i\omega)\;E +i\nabla^2 E +  P \; f(|E|^2)\; E \nonumber \\ 
&-& \Gamma(\omega+\epsilon \nabla^2)^2 E \, ,
\end{eqnarray}
\normalsize
where $E$ is the complex field, $E_{IN}$ is the (real) amplitude of the external forcing, $\omega$ is the frequency difference between the unperturbed field and the external driver, $\nabla^2$ is the transverse Laplacian, $P$ is the laser pump, $f(|E|^2)$ is $1-|E|^2/3$ for the  laser \cite{Mayol02} and $sinc^2(|E|)$  for the optical parametric oscillator \cite{Oppo13}. Time is normalised to the photon decay rate in the optical cavity and space to $\sqrt{L \lambda / 4 \pi}$ where $L$ is the cavity length and $\lambda$ the wavelength. Finally, $\Gamma$ is zero for the FCGL model and one for the FCSH case, where $\epsilon$ is a small parameter (here fixed at 0.01) due to the fast dynamics of atomic variables in lasers \cite{Lega94}. The FCGL and FCSH models can be applied in many other systems, e.g. chemical oscillations \cite{Hemming02}, granular media \cite{Yochelis06} and hydrodynamics \cite{Elphick97}.

\begin{figure}[h]
\centering
\includegraphics[width=0.95\linewidth]{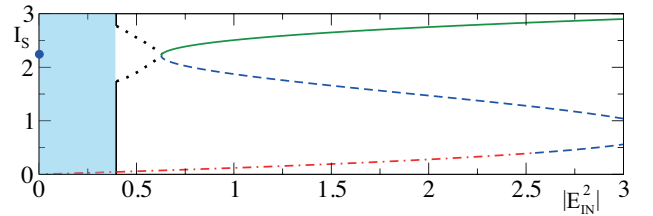}
\caption{(Color online) Stationary intensity of plane waves and their stability (stable = solid green, unstable with real eigenvalues = dashed blue, unstable with complex eigenvalues =  dash-dotted red) for the FCGL model with $P=4$ and $\omega=0.53$. The black dotted lines represent the minima and maxima of stationary hexagonal patterns, the vertical lines where the optical turbulent state starts (shaded area). The blue circle is the stationary intensity of the laser with no injection. The FCSH model displays very similar results.}
\label{fig1}
\end{figure}

\begin{figure}[h]
\includegraphics[width=0.85\linewidth]{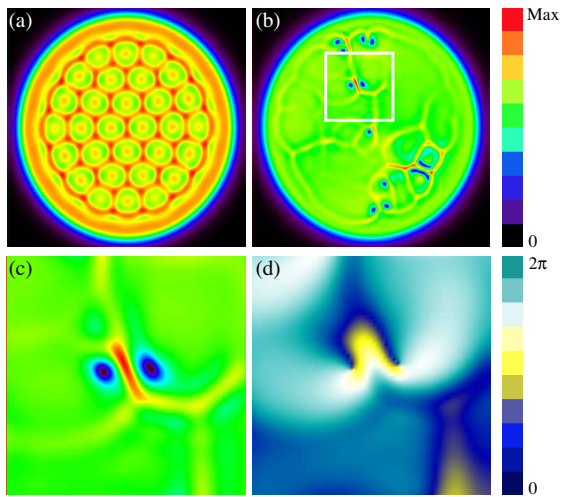}
\caption{(Color online) Transverse intensity for (a) hexagonal Turing pattern and (b)-(c) optical vortex-mediated turbulence. Corresponding transverse phase distribution of two oppositely charged vortices (d). Simulations of the FCGL equation with parameters, $P=6, \omega=0.77, E_{IN}=1.00$ (a) and $E_{IN}=0.95$ (b)-(d), see \cite{SuppMat}. The beam radius is 10$\pi$. (c) and (d) correspond to the area of the white square in (b).}
\label{fig2}
\end{figure}

\begin{figure}[h]
\includegraphics[width=0.95\linewidth]{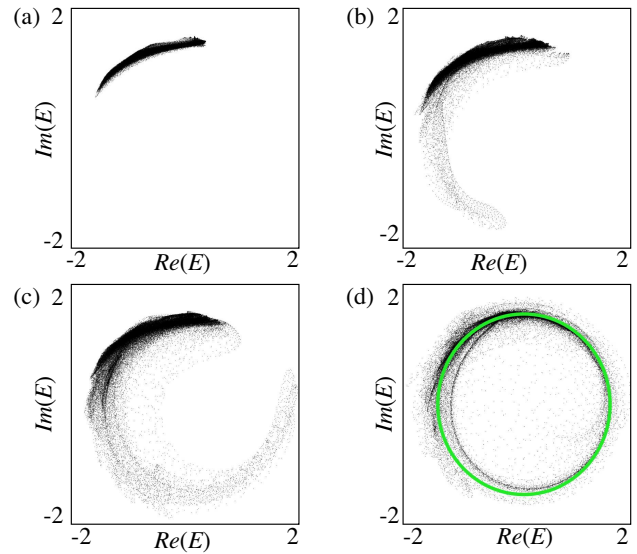}
\caption{(Color online) Field distributions in the Argand plane of the unstable hexagonal pattern at t=0 (a), during phase instability (t=183) (b), in a regime of amplitude instability (t=236) (c) and in a turbulent state (t=472). The green circle in (d) is the Adler  limit cycle. Simulations of the FCGL equation with the same parameters as Fig. \ref{fig1} and $E_{IN}=0.60$.}
\label{fig3}
\end{figure}

The cases of relevance are obtained when the detuning $\omega$ is different from zero. In this case the frequency locked states that one observes at large driving amplitudes become unstable upon decreasing the driving $E_{IN}$. For fixed values of $\omega$ and $P$, the homogeneous stationary states of Eq. (\ref{model}) have a typical $S$-shaped dependence on $E_{IN}^2$ as displayed, for example, in Fig. \ref{fig1} where the stability of these solutions to perturbations of zero wave-vectors is shown. 
The uppermost lines in the $S$-shaped curves of Fig. \ref{fig1} correspond to the homogeneous locked states where the external driving is large enough to overcome the frequency difference with the injected device. When increasing $E_{IN}$, a saddle-node bifurcation heralds the onset of the frequency and phase locked homogeneous states. When, instead, decreasing the parameter $E_{IN}$ the homogeneous solution loses stability to spatially periodic patterns with a critical wave-vector given by $k_c=\sqrt{\omega}$. In Fig. \ref{fig1} the maximum and minimum intensities of the hexagonal patterns obtained numerically \cite{SuppMat} when reducing the external driver are displayed via a black dotted line. Although the phase of the pattern is periodically modulated in space, the stationary character of these pattern solutions demonstrates that they are locked to the frequency of the injection. A typical hexagonal structure in the case of the FCGL equation with finite size input beams is shown in Fig. \ref{fig2}a. Note that all the results presented in this paper remain valid in the limit of transverse periodic boundary conditions.
As the amplitude $E_{IN}$ of the external drive is further reduced (see shaded area in Fig. \ref{fig1}), spatially periodic patterns become unstable and a regime of unlocked dynamics sets in \cite{Coullet92,Oppo13}. Figure \ref{fig3} shows the temporal evolution of an unstable hexagonal pattern in an Argand ($Im(E)$ versus $Re(E)$) diagram. The hexagonal pattern (see Fig. \ref{fig3}a) is a phase bound solution that progressively loses stability along a circle in the Argand diagram corresponding to a phase instability (Fig. \ref{fig3}b). This phase instability then grows into an amplitude instability (Fig. \ref{fig3}c) that leads to the formation and annihilation of pairs of oppositely charged vortices (see Fig. \ref{fig2}c and d) and a regime of spatio-temporal irregularity similar to the vortex-mediated turbulence described in \cite{Coullet89} in the CGL in the absence of forcing (see Fig. \ref{fig3}d). The helical waves propagating around the defects act as the driving force behind the turbulent state. A typical instantaneous intensity distribution of this turbulent state is presented in Fig. \ref{fig2}b \& c. The interacting vortices correspond to the localized regions of zero amplitude (shown in black). Note that the turbulent dynamics of vortices is deterministically driven by the spatially coupled nonlinearity and not by the superposition of random waves typical of optical speckles \cite{Shvartsman94}. Indeed, in the case of speckle, the field distribution in the Argand plane has a Gaussian shape as opposed to the almost circular one shown in Fig. \ref{fig3}d. There are also noticeable intrinsic differences in the field correlations \cite{Berry00} and in the PDFs of the intensity (see below). 

\begin{figure}[h]
\includegraphics[width=0.95\linewidth]{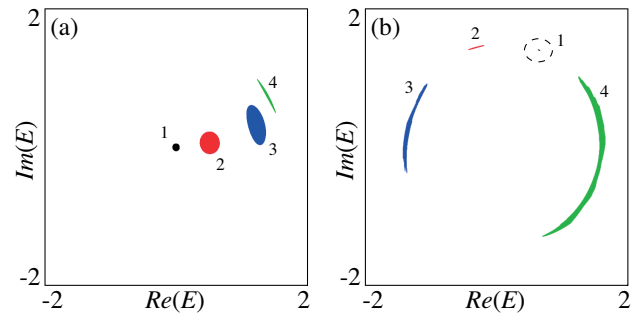}
\caption{(Color online) Field distributions in the Argand plane for the FCGL equation with the same parameters as Fig. \ref{fig1} and $E_{IN}=0.70$. (a): t=0 (black), t=0.40 (red), t=0.75 (blue) and t=1.4 (green). (b): t=90 (black), t=256 (red), t=280 (blue) and t=314 (green).}
\label{fig4}
\end{figure}

To better understand the nature of the turbulent state in forced models (\ref{model}) we consider dynamical solutions corresponding to unlocked oscillations and their robustness to spatially dependent perturbations in the FCGL model ($\Gamma=0$) where analytical predictions are feasible. In the absence of spatial coupling an approximate limit cycle trajectory for the field can be found by period averaging methods, $\mathrm {E=A_0 \left[ \cos(\phi(t)) + i  \sin(\phi(t)) \right]}$, \cite{Mayol02,Erneux10}, 
%
where $A_0^2=3(P-1)/P$, $\phi(t)$ is well approximated by its period average $\Omega t+\pi$ with $\Omega=\sqrt{\omega^2-\omega_L^2}$ and $\omega_L=E_{IN}/A_0$. When $\Omega$ is real, i.e. in the absence of locked states, the trajectory 
is the phase-drift solution of the Adler equation \cite{Adler46}, $d_t \phi = \omega -\omega_L \sin(\phi(t))$.
%
%
Such solution is clearly phase unbound and is superimposed onto the Argand diagram in the turbulent regime of Fig. \ref{fig3}d to show that its underlying dynamics is ruled by the unlocked state. The accuracy of the approximate solution 
has been checked for a wide range of $E_{IN}$ values in the FCGL model. The excursions in intensity do not exceed 10\% while those in frequency are well within 1\%. We have then proceeded to study the stability of the spatially synchronized oscillation by including spatial coupling in the FCGL. The stability eigenvalues of the spatially synchronized limit cycle 
are given by 
%
\begin{equation}
\lambda_{\pm} = - (P-1) \pm \sqrt{ (P-1)^2 - (\omega-k^2)^2}
\label{LSA}
\end{equation}
where $k$ is the spatial wave-vector. At the critical wave-vector for pattern formation $k_c=\sqrt{\omega}$, the stability $\lambda_+$ is marginal but perturbations due to the approximate nature of 
$E$ 
induce a slow instability of the synchronous oscillation. The eigenvector associated with $\lambda_+$ is along the limit cycle, again demonstrating a phase instability. As mentioned earlier, this phase instability grows into an amplitude instability and then into vortex-mediated turbulence as demonstrated numerically in Fig. \ref{fig4}, starting from low amplitude noise. A homogeneous zero state with added noise quickly evolves towards the unstable limit cycle
(from 0 to 4 in Fig. \ref{fig4}a). The limit cycle dynamics first synchronises the spatial oscillations (see the narrow line at $t=90$ in Fig. \ref{fig4}b) and then moves towards the vortex turbulence state via phase (Fig. \ref{fig4}b) and amplitude (see  Fig. \ref{fig3}c) instabilities. We outline that the mechanism of spontaneous vortex creation in the FCGL and FCSH is not trivial. In contrast with the CGL, stationary vortex solutions are not possible in driven systems like (\ref{model}) as all locked states have bound phases around that of the injection. However, at low driving amplitudes, moving vortices and vortex-mediated turbulence in (\ref{model}) are possible due to the Adler unlocked dynamics of the limit cycle trajectory. 
It is known \cite{Prati10} that the adiabatic elimination of the polarization variable introduces an all wave-vector instability of the spatially homogeneous state below the point where the linear stability of the lower branch of the $S$-shaped homogeneous state predicts complex conjugate eigenvalues (see Fig. \ref{fig1}). This feature, in principle, may have serious consequences in the turbulent regimes. A second important consequence of our analysis, however, is that Eq. (\ref{model}) for $\Gamma=0$ displays a very fast dynamics that takes the system towards the limit cycle  
where large wave-vector instabilities are promptly eliminated (see  
Fig. \ref{fig4})). The large wave-vector instability of the lower branch of the homogeneous stationary states is not present in the case of FCSH when $\Gamma=1$.

\begin{figure}[h]
\centering
\includegraphics[width=0.95\linewidth]{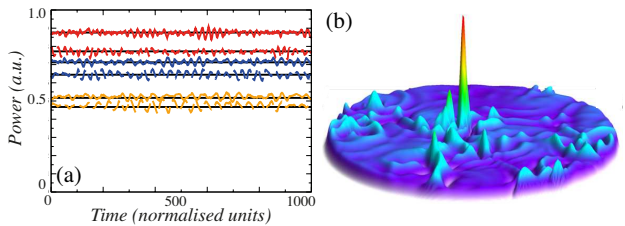}
\caption{(Color online) (a) Time evolution of the power for three values of the pump $(P=2,4$ and $5)$ and for the FCGL (solid lines) and FCSH (dashed lines) models during vortex turbulence. The horizontal lines are the predictions from the approximate Eq. (\ref{eqn:power}). (b) RW spike in the transverse intensity for the FCGL with $P=6,  \omega=0.77, E_{IN}=1.00$. The peak intensity is 42.22, the average is 4.93 and the standard deviation is 1.57 (all in normalised units).}
\label{fig5}
\end{figure}

This mechanism for vortex turbulence is essential for the generation of RWs in externally driven systems described by the spatio-temporal dynamics of Eq. (\ref{model}). These systems are outside thermodynamic equilibrium, do not display relaxation oscillations and present a delicate balance between the energy input and the losses (the pump $P$, injection $E_{IN}$, nonlinearity $f(|E|^2)$ and output mirror). During the turbulent evolution, the total power ${\cal P}$ remains almost constant at values close to those of the laser with no injection (see Fig. \ref{fig5}a). By considering the energy density and the energy flux of the FCGL equation \cite{Akhmediev05} the time evolution of the power is given by
\begin{equation}
\frac{\partial \cal P} {\partial t} = 2 \int \left[ E_{IN} Re(E) + \left( P - 1  - \frac{P}{3} |E|^2 \right) |E|^2 \right] dx dy
\label{eqn:power}
\end{equation}
where $(x,y)$ is the transverse plane. For the approximate limit cycle solution
the power ${\cal P}$ is conserved at the value of $\pi w_0^2 A_0^2$ where $w_0$ is the beam width of the input laser. In the turbulent state, however, maintaining an almost constant power in the presence of moving vortices of zero intensity implies the simultaneous appearance of large amplitude spikes. If the vortex density is large, multi-vortex collisions can occur with the production of large, short-lived spikes in the intensity (see Fig. \ref{fig5}b). Short-lived large intensity spikes are rare but possible events, fitting the characteristics of RWs. The particular shape and symmetry of these spikes is crucially determined by the number and position of the surrounding vortices, a feature that is unique to this particular mechanism of RW formation. RWs in single transverse mode class-B lasers with injected signals have been observed in \cite{Bonatto11} but due to relaxation oscillations and not to 
2D vortex turbulence. In fact, without spatial coupling due to diffraction, no RWs can be observed in systems described by Eq. (\ref{model}). To characterize our spatio-temporal RWs, we use a commonly accepted definition of statistically rare events \cite{Bonatto11,Dudley14,Oppo13}, i.e. if the intensity of the field at a spatial point over a long period of time is greater than the mean wave height plus eight standard deviations then the wave can be classified as an extreme event or RW, similarly to the significant peak intensity method  \cite{Dudley14}. 

\begin{figure}[h]
\centering
\includegraphics[width=0.95\linewidth]{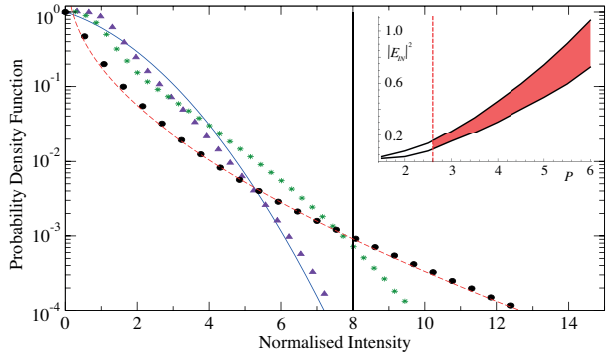}
\caption{(Color online) Intensity PDFs for the FCGL model: $P=2, \omega=0.3$, $E_{IN}=0.24$ (purple triangles) with Gaussian fit (blue line) and $P=8, \omega=2.4$, $E_{IN}=3.40$ (green stars) and the $sinc^2$ model: $P=8,  \omega=1$, $E_{IN}=1.48$ (black circles) with Weibull fit (red dashed line) \cite{SuppMat}. The vertical black line is the threshold for defining waves as an extreme event. 
Inset: Parameter region for RWs (red area) in the FCGL. The upper curve is the 
pattern to vortex-mediated turbulence transition, the lower curve the turbulence to target pattern transition. No RWs are observed below the red dashed line.}
\label{fig6}
\end{figure}

In Fig. \ref{fig6} we show PDFs for different regimes of vortex turbulence.
The wave statistics is well-approximated by a Gaussian fit when the pump intensity is low (blue solid line in Fig. \ref{fig6}). 
At higher pump intensities  the long-tailed PDFs show mass generation
of RWs. Here the statistics changes drastically and is very well approximated by a Weibull distribution \cite{Onorato13, SuppMat} (red dashed line in Fig. \ref{fig6}).
Note that non-Gaussian PDFs cannot be replicated by superpositions of random waves. We also note that the RWs in vortex turbulence demonstrated in Fig. \ref{fig6} are different from those due to vorticity in models of inviscid fluids \cite{Abrashkin13}. The CGL and CSH have been shown to be equivalent to the flow of a compressible and viscous fluid with density $\rho=|E|^2$ and velocity {\bf v}=$\nabla \phi$ where $\phi$ is the phase of the field \cite{Brambilla91}. In the case of our forced systems, $\nabla \times${\bf v} remains extremely close to zero in the locations where RWs are observed. We conclude that our RWs are due to the interaction of free vortices in the absence of vorticity.

Finally, we show in the Fig. \ref{fig6} inset the wide parameter region where we observe RWs induced by vortex turbulence in systems with external driving of the FCGL kind. Very similar results have been obtained for the FCSH and different nonlinearities such as those of the FCSH and the singly resonant OPO, thus demonstrating the universality of the phenomenon.


In conclusion, we demonstrate a mechanism for producing RWs in the transverse area of externally driven nonlinear optical devices via vortex turbulence. Given the universality of our model, this mechanism should be observable in a large variety of systems. Models of lasers with injected signal, where the invariance of the Adler limit cycle is well known \cite{Oppo86,Mayol02}, can be easily extended to semiconductor media \cite{Gustave15} and to class B lasers, thus including the largest majority of solid state lasers. 
Outside optics, vortex-mediated turbulence without driving has been observed in nematic liquid crystals \cite{Frisch95}, chemical reactions \cite{Ouyang96} and fluid dynamics \cite{Bodenschatz00}. 
In the unlocked regime of these systems with driving, vortex turbulence can excite RWs and lead to the formation of highly inhomogeneous fields with non-Gaussian statistics. 

The prototype model used to describe RWs is the Non-Linear Schr\"{o}dinger (NLS) equation \cite{Onorato13,Solli07,Dudley14}. The FCGL and FCSH models studied here are active, non-conservative systems outside thermodynamic equilibrium where many of the methods developed for the NLS cannot be applied. In the NLS equation, as well as in the CGL and CSH equations, stationary vortex solutions are possible although mainly unstable. In the presence of forcing, vortices can only exist in dynamical states. It is exactly in these situations that we have demonstrated RWs close to regions of interaction of turbulent vortices. Because of universality, suitably perturbed NLS models may also display these features.

Unlike RWs in the longitudinal direction \cite{Dudley14}, the aspect ratios required for transverse RWs induced by 2D vortex turbulence are extremely small (typical input beams have diameters less than 1$mm$) and the statistics require times of the order of hundreds of $\mu s$. The small aspect ratio, the full 2D character and the quick dynamics represent the major advantages of transverse optical devices in studying the generation and control of RWs with applications, by universality, in hydrodynamics and oceanography. 

\section{Acknowledgments}
CJG acknowledges support from an EPSRC DTA grant EP/M506643/1.


\end{document}